# Strongly interaction-enhanced valley magnetic response in monolayer WSe$_2$

Zefang Wang, Kin Fai Mak*, and Jie Shan*

Department of Physics and Center for 2-Dimensional and Layered Materials,
The Pennsylvania State University, University Park, Pennsylvania 16802-6300, USA
*E-mails: kzm11@psu.edu; jus59@psu.edu

**Coulomb interaction between electrons lies at the heart of magnetism in solids [1,2]. In contrast to conventional two-dimensional (2D) systems, electrons in monolayer transition metal dichalcogenides (TMDs) possess coupled spin and valley degrees of freedom by the spin-orbit interaction [3,4]. The electrons are also strongly interacting even in the high-density regime because of the weak dielectric screening in two dimensions and a large band mass [5,6]. The combination of these properties presents a unique platform for exploring spin and valley magnetism in 2D electron liquids. Here we report an observation by magneto-photoluminescence spectroscopy of a nonlinear valley Zeeman effect, correlated with an over fourfold enhancement in the exciton g-factor in monolayer WSe$_2$. The effect occurs when the Fermi level crosses the spin-split upper conduction band, corresponding to a change of the spin-valley degeneracy from 2 to 4. The enhancement increases, shows no sign of saturation as the sample temperature decreases. Our result suggests the possibility of rich many-body ground states in monolayer TMDs with multiple internal degrees of freedom.**

Electrons in monolayer transition metal dichalcogenide (TMD) semiconductors with a honeycomb lattice structure possess a two-fold valley degree of freedom, corresponding to the K and K' point of the Brillouin zone [3-5]. Because of the strong spin-orbit interaction, the bands are spin split with the valley and spin locked to satisfy the time reversal symmetry [3-5] (Fig. 1a). Similar to the spin, the valley carries a magnetic moment and has been proposed as a new type of information carriers [3,7]. Several new valley dependent phenomena, including the valley contrasting optical selection rules [8-12], valley Zeeman effect [13-18] and valley Hall effect [19,20], have emerged in the independent-particle picture and provided means to manipulate the valley polarization. In particular, the valley exciton splitting in an out-of-plane magnetic field has been shown to depend linearly on the field up to 65 Tesla [18] and an exciton g-factor between -2 and -4 has been reported for various monolayer TMDs [6, 13-18]. On the other hand, even in the relatively high-density regime ($\sim 5 \times 10^{12}$ cm$^{-2}$), electrons in monolayer TMDs are strongly interacting with the Coulomb energy (~ 100's meV) dominating all other energy scales (Fermi energy, conduction band spin splitting at K/K' ~ 10's meV for WSe$_2$) [6]. The valley magnetic response of the independent-particle picture is thus expected to be modified by the strong electron-electron interaction and the system may even develop a magnetically ordered ground state [21-24]. A unique scenario emerges when the Fermi level crosses the spin-split upper conduction band (Fig. 1a), where the spin-valley degeneracy $l_s l_v$ changes from 2 to 4 ($l_s$ and $l_v$ stand for the spin and valley degeneracy, respectively) [6]. We observe a strongly enhanced valley magnetic response, which can be understood as a consequence of a step rise in the exchange interaction strength that is highly sensitive to the number of internal degrees of freedom [21, 22, 24]. Our result opens up new possibilities of exploring strongly interacting electron systems with multiple internal degrees of freedom beyond the conventional multi-valley semiconductor quantum wells of Si [25-29] and AlAs [30, 31].

We examine the valley Zeeman effect in monolayer WSe$_2$ over a wide electron doping range using dual-gate field-effect devices. Monolayer WSe$_2$ is embedded in hexagonal boron nitride (hBN) substrates with few-layer graphene as both contact and gate electrodes. As reported earlier, encapsulation of the material in hBN produces high-quality samples [6, 32, 33], and the use of dual local gates enables high doping densities. Figure 1b shows a contour plot of the photoluminescence (PL) spectra of monolayer WSe$_2$ at varying electron densities $n$ at 5 K under zero magnetic field. The corresponding doping dependence of the integrated PL intensity is shown in Fig. 1c. The doping density was calibrated using a capacitance model and the gate voltages. (see ref. [6, 34]). A sharp emission feature with a linewidth $\Gamma \sim 3 - 7$ meV is observed. It arises from optical transitions between the upper conduction and upper valence band that have the same spin [4, 6, 34] (modified by the electron-hole and electron-electron interaction [5, 35, 36]) (Fig. 1a). The emission peak red shifts with doping because of the combined excitonic and band gap renormalization effects [34-37]. Doping into the upper conduction band at $n_0 \approx 6.4 \times 10^{12}$ cm$^{-2}$ (dashed line, Fig. 1c) causes a sharp increase in the PL intensity because the occupancy of the conduction band and the radiative recombination rate drastically increase [34]. The value $n_0$ can also be determined from the Pauli blocking effect on the optical absorption [6, 34]. We include in the top axis of Fig. 1c the Wigner-Seitz radius $r_S = \frac{1}{\sqrt{\pi n} a_B^*}$ to measure the interaction strength. Here $a_B^* = \epsilon \frac{m_0}{m_c} a_B$ is the Bohr radius $a_B$ modified by the average background dielectric constant of hBN $\epsilon = \sqrt{\epsilon_\perp \epsilon_\parallel} \approx 4.15$ ($\epsilon_\perp \approx 2.5$ [6] and $\epsilon_\parallel \approx 6.9$ [38] are the out-of-plane and in-plane dielectric constant of hBN, respectively), and the conduction band mass is in units of the free electron mass ($m_c/m_0 \approx 0.4$ [39]). The large $r_S$ (3.3 – 7.0) suggests that the system is in the strongly interacting regime. More details on the device fabrication and characterization are provided in Methods and Supplementary Sect. 1.

We measure the exciton valley Zeeman splitting ($E_Z$) at different doping densities by magneto-PL spectroscopy. The left- and right-handed PL spectra under linearly polarized excitation and an out-of-plane magnetic field $H$ are recorded. They correspond to the transitions at the K' and K valleys, respectively, according to the valley contrasting optical selection rules [3]. The PL spectra rather than the reflectance contrast spectra were measured for higher precision in $E_Z$ since they are background-free. Figure 1d illustrates the results under -2 T for two representative electron densities $5.0 \times 10^{12}$ ($< n_0$) and $8.6 \times 10^{12}$ cm$^{-2}$ ($> n_0$). A clear increase in the splitting and hence a larger valley magnetic response are observed for the latter. We determine $E_Z$ as the difference between the weighted peak energy of the right- and left-handed PL spectra. The uncertainty for each peak value is estimated as $\Gamma/\sqrt{N}$, where $N$ is the total PL count (which is $\sim 5 \times 10^5$ for a typical integration time of 3 s for our highly luminescent samples). Figure 1e shows the results for $E_Z$ under $H = 0$ and -2 T. The uncertainty in $E_Z$ is generally about $5 - 30$ $\mu$eV, which is much smaller than the emission linewidth (i.e. super-spectral-resolution). A small offset (<50 $\mu$eV) is present at $H = 0$, which is likely due to systematic errors in the selection of the light polarization. The sensitivity of our measurement at the peak response is ~ 10 mT. More details on the magneto-PL spectroscopy and analysis of the valley Zeeman splitting are provided in Methods and Supplementary Sect. 2.

The magnetic field dependence of $E_Z$ at 5 K is shown in Fig. 2a for several representative doping densities. Before doping into the upper conduction band ($n = 4\times10^{12}$ and $5\times10^{12}$ cm$^{-2}$), the dependences are linear for the field ranging from -8 T to 8 T, as reported earlier [13-18]. After doping into the upper conduction band for $n =$ $6.7\times10^{12}$, $7.0\times10^{12}$ and $7.3\times10^{12}$ cm$^{-2}$, the dependences become nonlinear with $E_Z$ saturated around a critical field $H^*$. The critical field increases with doping density. With a further increase of $n$ (= $8.3\times10^{12}$), $H^*$ is outside the field range and the dependence appears linear again, but with a significantly larger slope. We extract the exciton g-factor (Fig. 2b) from the slope of the $E_Z - H$ dependence $g_X = \frac{1}{\mu_B}\frac{dE_Z}{dH}$ at $H = 0$, where $\mu_B \approx$ 0.0579 meV/T is the Bohr magneton. We also estimate $H^*$ (Fig. 2c) by fitting the phenomenological expression $E_Z = g_X\mu_B H^* \tanh(H/H^*)$ to the $E_Z - H$ dependences (dashed lines, Fig. 2a). For low doping levels ($n < n_0$), $|g_X|$ decreases slowly with $n$ from ~ 5 to 3, which is consistent with the reported values for monolayer WSe$_2$ [15-17]. The variation in the reported value may originate from the different doping levels present in unintentionally doped WSe$_2$ samples. As $n$ increases beyond $n_0$, $|g_X|$ increases rapidly to ~12, and then decreases with a further increase of $n$. The strong enhancement in $|g_X|$ is correlated with the doping dependence of $H^*$ shown in Fig. 2c. The critical field increases from ~ 4 T near $n_0$, followed shortly by a linear dependence on density $\propto (n - n_0)$ (solid line, Fig. 2c).

The observed nonlinear field dependence of the valley Zeeman splitting and the strong enhancement of the exciton g-factor are unexpected in the absence of interaction effects. In the independent-particle picture, the exciton Zeeman splitting is given by the difference between the Zeeman splitting of the upper conduction band and of the upper valence band [13-18]. The total magnetic moment of a charge carrier of a particular band consists of the atomic orbital, spin, and inter-atomic orbital (or valley) contributions [14, 15]. In the two-band $k \cdot p$ model, the atomic orbital moments are $\pm 2\mu_B$ for the valence bands and 0 for the conduction bands near the K/K' point, reflecting the properties of the $d$-orbitals of the W atom that form the bands [3]. The spin contribution to $E_Z$ is largely canceled since optical transitions are allowed only for bands of the same spin [14, 15]. Finally, the inter-atomic orbital moments are $\pm \frac{m_0}{m_{v/c}}\mu_B$ for the K/K' point [14, 15], where the Bohr magneton is modified by the effective mass $m_c$ for the conduction band and $m_v$ for the valence band. The exciton g-factor is thus given by $g_X = -4 - 2\left(\frac{m_0}{m_v} - \frac{m_0}{m_c}\right)$. In case of similar conduction and valence band masses such as in monolayer WSe$_2$, we obtain $g_X \approx -4$ and a linear valley Zeeman effect $E_Z \approx -4\mu_B H$, independent of doping density.

Our experimental observations can be understood by considering the strong Coulomb interaction. In the vicinity of $n_0$, $r_s$ (which measures the average inter-particle separation in units of the effective Bohr radius, $\approx 4.2$) is nearly a constant (Fig. 1c), but the spin-valley degeneracy $l_s l_v$ increases from 2 to 4 when the Fermi level crosses the upper conduction band. A qualitative measure of the exchange interaction strength is provided by $\sim \sqrt{l_s l_v} r_S$ [24], which has a sudden increase over a narrow doping range when the number of electron species doubles. Our results suggest that in this system the increased exchange interaction and exchange field ($H_{ex}$) favor larger spin and valley polarization (i.e. a larger Zeeman splitting and g-factor) to lower the total energy of the system. The effect of $H_{ex}$ remains significant until $H$ reaches $H^*$, at which the electrons in the upper conduction band become fully spin and valley polarized. Beyond $H^*$, the effect of $H_{ex}$ diminishes, which explains the saturated $E_Z - H$ dependences (Fig. 2a).

Figure 2d illustrates schematically the Zeeman effect for the spin-split conduction bands near the K/K' point for $n < n_0$ (left) and $n > n_0$ (right). Under an external magnetic field ($H < 0$), the upper conduction band down (up) shifts for the K (K') valley since the valley Zeeman and spin Zeeman shift are in the same direction [15]. (The shift direction of the lower conduction band is determined by the relative importance of these two effects.) The Zeeman shift of the bands is larger for $n > n_0$ because of the enhanced exchange field. In particular, the right diagram shows the case of $H = H^*$, at which the Fermi level lies at the bottom of the upper conduction band at the K' point and the electrons in the upper conduction band are fully spin and valley polarized. We can relate the critical field $H^*$ to the doping density $n$ as $H^* = \frac{\hbar^2 \pi (n - n_0)}{m_c |g_c| \mu_B}$ (ref. [26, 27]), where $g_c$ is the upper conduction band g-factor and identical masses are assumed for the upper and lower conduction bands. A comparison of this simple picture with the experimental data (solid line, Fig. 2c) yields $m_c |g_c| \approx (4.7 \pm 0.1) m_0$. The discrepancy at low doping densities is likely due to Fermi level broadening by impurities/defects, which also broadens the enhancement in $|g_X|$ in Fig. 2b. In the independent-particle picture, the conduction band g-factor is given by $|g_c| = 2 + 2 \frac{m_0}{m_c}$ [15], where the two terms correspond to the spin and valley contribution, respectively (the atomic orbital contribution = 0 for the conduction bands). Using $m_c \approx 0.4\, m_0$ [39], we estimate the "non-interacting" value of $m_c |g_c| \approx 2.8\, m_0$, which is much smaller than the experimental value, in support of the interpretation of an interaction-enhanced magnetic response. Enhancements in $m_c |g_c|$ in similar material systems but under different conditions have also been reported by recent studies [40-42].

We make several comments regarding the results above. First, the enhancement in the electron magnetic susceptibility ($\sim m_c |g_c|$) near $n_0$ is likely dominated by the g-factor enhancement. This is supported by ref. [6], which shows only a weak enhancement in $m_c$ as $n$ increases through $n_0$. This behavior is distinct from that in Si quantum wells [25-29], where the mass renormalization also plays an important role. Further theoretical studies on the origin of the magnetic susceptibility enhancement are warranted. Second, the weaker variation in $|g_X|$ away from $n_0$ (both below and above) is likely caused by a changing $r_S$. As doping increases, $r_S$ decreases (i.e. interaction effects weaken) and a reduced $|g_X|$ is observed, in agreement with theoretical calculations [23, 24]. Last, the over fourfold enhancement in $|g_X|$ observed in monolayer WSe$_2$ is much larger than that has been reported (less than twofold) in multi-valley Si quantum wells [25, 28, 29]. The maximum exciton g-factor observed in this study (~ 12) is close to 40% of the value obtained for monolayer WSe$_2$ proximity coupled to a ferromagnetic insulator [43], which further illustrates the importance of the exchange interaction above $n_0$.

Finally, we briefly discuss the temperature dependence of $|g_X|$. Figure 3a shows the density dependence of $|g_X|$ determined from the valley Zeeman splitting at 2 T at temperatures ranging from 10 to 80 K. The strong enhancement in $|g_X|$ for $n > n_0$ emerges only at low temperatures (< 40 K). Figure 3b shows the temperature dependence of $|g_X|$ at two representative doping densities. For $n = 4.0 \times 10^{12} < n_0$, $|g_X|(\approx 4)$ is nearly temperature independent. On the other hand, for $n$ slightly above $n_0$, $|g_X|$ grows significantly as temperature decreases. No sign of saturation is observed down to 5 K. The result shows that in comparison with $\frac{\hbar^2 \pi (n - n_0)}{2 m_c}$, the thermal broadening of the Fermi level has to be small in order to reveal the interaction-enhanced magnetic response. Moreover, the absence of saturation in the enhancement of $|g_X|$ suggests that the magnetic response of the system could be further enhanced at lower temperatures and in higher quality samples. Our study has paved the path for the search of the interaction driven ferromagnetic instability in monolayer TMDs with unique electronic structures and multiple internal quantum degrees of freedom.

**Methods**

**Device fabrication**
Dual-gate field-effect devices of monolayer WSe$_2$ were fabricated using the mechanical exfoliation and dry transfer method. Details have been reported elsewhere [6, 32]. In brief, few-layer graphene, thin hexagonal boron nitride (hBN) flakes of about 20-nm thickness and monolayer WSe$_2$ were first exfoliated from bulk crystals onto SiO$_2$/Si substrates. They were then picked up layer-by-layer using a polymer stamp to form an hBN/WSe$_2$/hBN vertical stack with few-layer graphene as both contact and gate electrodes. The complete stack was then deposited onto SiO$_2$/Si substrates with pre-patterned Au electrodes so that the graphene electrodes were in contact with separate metal electrodes. In the measurements, the contact electrode was grounded and bias voltages were applied to the gate electrodes to vary the doping density in monolayer WSe$_2$. Image of the device is provided in Supplementary Sect. 1.

**Magneto-optical measurements**
Photoluminescence (PL) spectroscopy was performed in the Faraday geometry in an Attocube closed-cycle cryostat (attoDry1000) under varying magnetic fields, gate voltages and temperatures. A linearly polarized excitation beam centered at 532 nm was focused into a diffraction-limited spot on the sample by a microscope objective. The excitation power was kept around 100 $\mu$W to limit laser heating of the sample while high PL counts can still be obtained. The PL from the sample was collected by the same objective, passed through polarization selection optics, and detected by a spectrometer equipped with a charge-coupled-device (CCD). The typical integration time for each spectrum is 3 s. The left and right circularly polarized components of the PL were selected by a combination of a quarter-wave plate, half-wave plate and linear polarizer. The half-wave plate was mounted on a motorized rotator to rapidly switch the emission polarization from one handedness to the other. To minimize the long-term drift of the setup, the left- and right-handed PL spectra were collected within 10 s of each other.

**Analysis of the PL spectra**

Because of the complicated recombination process of the electron-hole pairs in the presence of a strongly interacting electron liquid, the PL spectra have an asymmetric line-shape, which cannot be described by a simple analytic function. To obtain the Zeeman splitting with high accuracy, we chose to determine the peak energy of the left and right circularly polarized emission by computing their statistically averaged peak energy instead of performing fits to the spectra. To focus on the main emission feature, we limited the spectral window of interest in which the PL intensity is above 20% of the peak intensity. The value of 20% was chosen to eliminate most contributions from the localized exciton emission. The average peak energy was determined as the weighted sum of the photon energy by the PL intensity for the chosen spectral window. All spectra were analyzed using the same procedure. Examples are provided in Supplementary Sect. 2.

**Figures and figure captions**

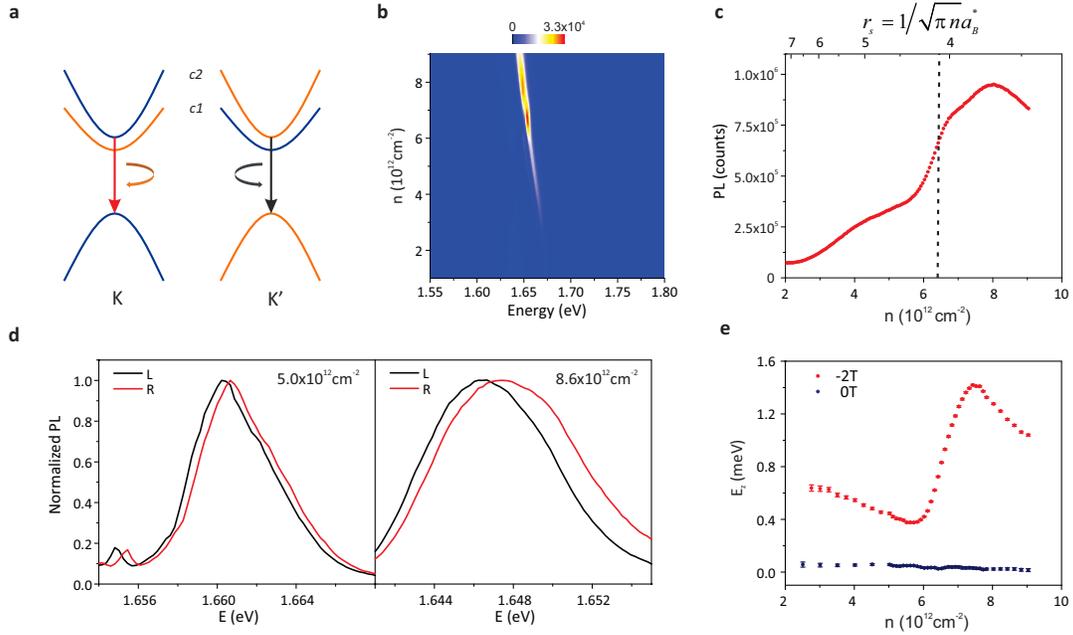

**Figure 1 | Magneto-photoluminescence of monolayer WSe$_2$. a,** Electronic band structure of monolayer WSe$_2$ at the K and K' valley including the two spin-split conduction bands (*c*1, *c*2) and the upper valence band. Optical transitions are allowed only between bands of the same spin (shown in the same color) in the same valley. The left and right circularly polarized light couple to the K' and K valley, respectively. **b,** Contour plot of the photoluminescence (PL) counts as a function of photon energy and electron doping density *n*. The integration time is 3 s. **c,** Spectrally integrated PL counts as a function of doping density *n* (bottom axis) and Wigner-Seitz radius $r_S$ (top axis). A sharp increase occurs around $n_0 \approx 6.4 \times 10^{12}$ cm$^{-2}$ (dashed line) when the Fermi level crosses band *c*2. **d,** Handedness resolved PL spectra (normalized) at doping densities $5.0 \times 10^{12}$ cm$^{-2}$ (left) and $8.6 \times 10^{12}$ cm$^{-2}$ (right) under -2 T. **e,** Zeeman splitting as a function of doping density under 0 T and -2 T.

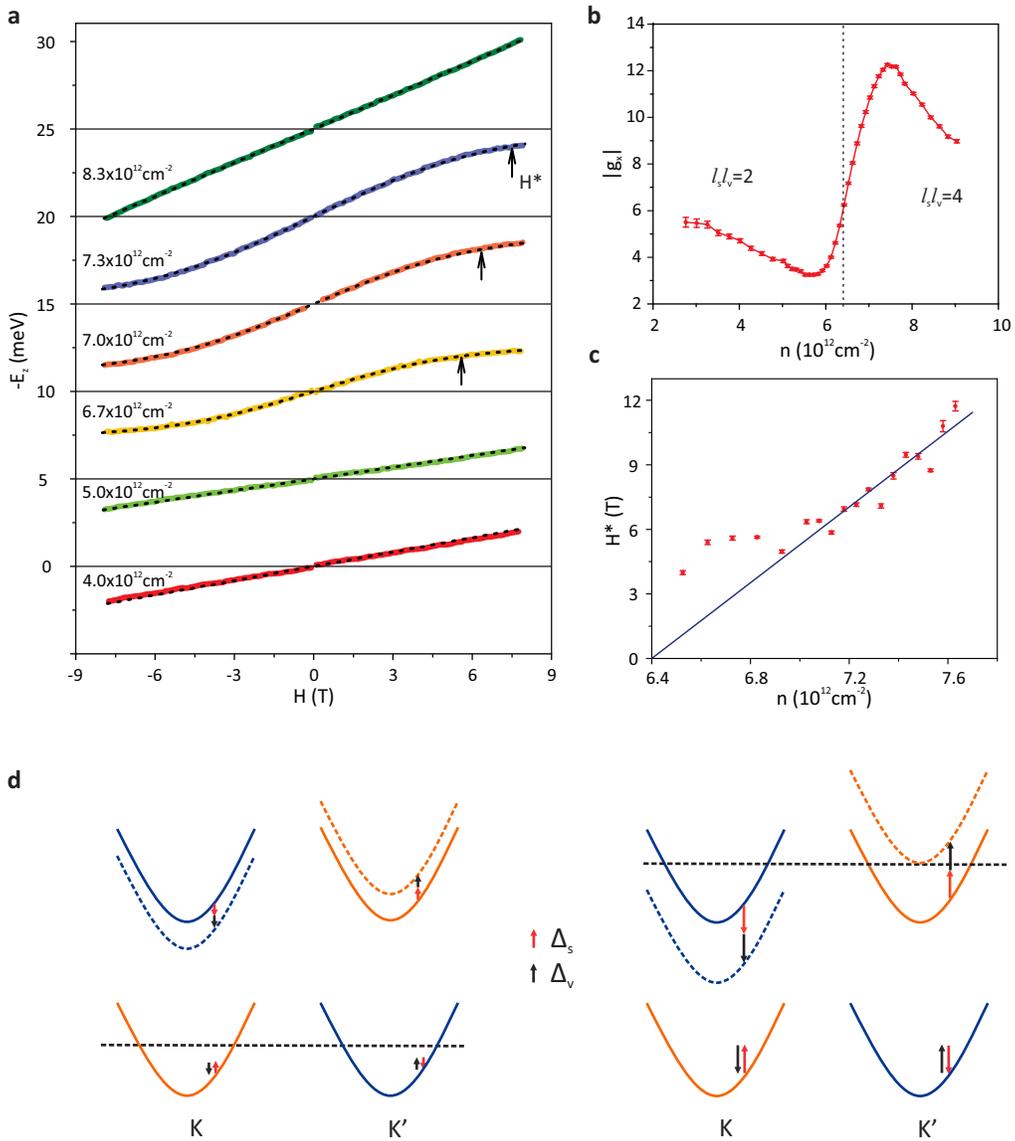

**Figure 2 | Interaction-enhanced valley Zeeman effect in monolayer WSe$_2$. a,** Valley Zeeman splitting as a function of magnetic field ranging from -8 T to 8 T at representative doping densities. The data at different $n$ are vertically shifted for clarity. Solid color lines are experimental data and dashed black lines are fits to the experimental data of the phenomenological function described in the text. The Zeeman splitting saturates at critical field $H^*$. **b,** Doping dependence of the exciton g-factor $|g_X|$ (symbols) determined from the slope of the field dependence of the Zeeman splitting at zero field. The error bars are estimated from the uncertainties of the PL peak energy of two different circular polarizations. Dashed line denotes $n = n_0$. **c,** Doping dependence of the critical field $H^*$ determined from the saturated Zeeman effect of (**a**). The solid line is a linear fit to the experimental data (symbols) with an x-intercept at $n_0$. **d,** Valley Zeeman shift ($\Delta_v$)

and spin Zeeman shift ($\Delta_s$) of the two spin-split conduction bands for $n < n_0$ (left) and $n > n_0$ (right) under magnetic field $H < 0$. The right diagram corresponds to the case of $H = H^*$, for which the upper conduction band is fully spin and valley polarized.

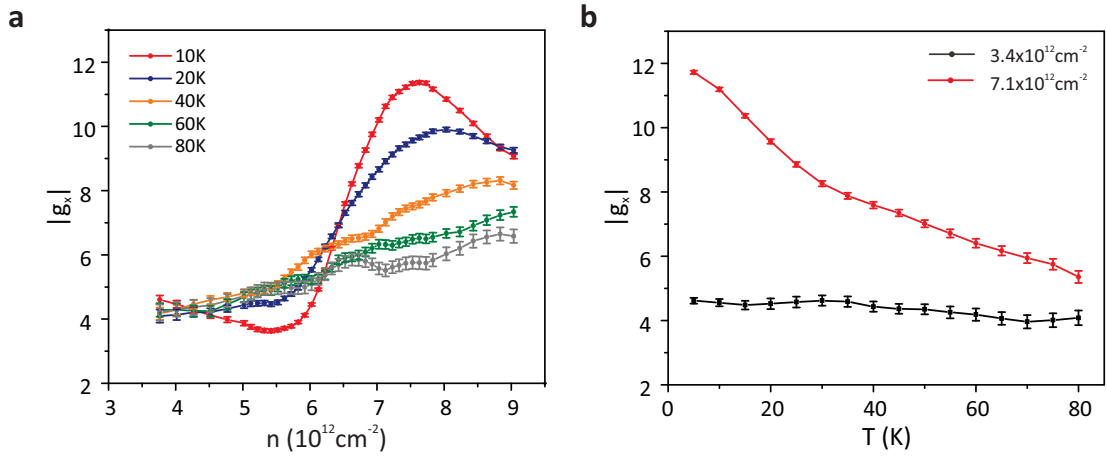

**Figure 3 | Temperature dependence of the valley Zeeman effect. a,** Exciton g-factor as a function of doping density at representative temperatures. **b,** Exciton g-factor as a function of temperature at representative doping densities. The g-factor values are determined from the Zeeman splitting measurement under 2 T.